\documentclass{PoS}
\PoS{PoS(LAT2005)271}
\usepackage{fontenc}
\usepackage{amssymb}
\usepackage{amsmath}
\usepackage{graphicx}
\usepackage{times}
\usepackage{mathptmx}
\usepackage{color}
\usepackage{ifpdf}

\ShortTitle{Perturbative Study of the Supersymmetric Lattice Model
from Matrix Model}
\title{Perturbative Study of the Supersymmetric Lattice Model
from Matrix Model}
\author{Tetsuya Onogi
, \speaker{Tomohisa Takimi} \\
Yukawa Institute for Theoretical Physics, Kyoto University,
Kitashirakawa-Oiwakecho, Sakyo, Kyoto 606-8502, Japan \\
E-mail: \email{onogi@yukawa.kyoto-u.ac.jp},
\email{takimi@yukawa.kyoto-u.ac.jp}}
\abstract{
We study the lattice model for the supersymmetric Yang-Mills theory 
in two dimensions proposed by Cohen, Kaplan, Katz, and Unsal. 
We re-examine the formal proof for the absence of susy breaking counter 
terms as well as the stability of the vacuum by an explicit 
perturbative calculation for the case of $U(2)$ gauge group.
Introducing fermion masses and treating the bosonic zero momentum mode 
nonperturbatively, we avoid the infra-red divergences in the
perturbative calculation. As a result, we find that there 
appear mass counter terms for finite volume which vanish
in the infinite volume limit so that the theory needs no fine-tuning. 
}
\FullConference{XXIIIrd International Symposium on Lattice Field
Theory \\ 
25-30 July 2005 \\
Trinity College, Dublin, Ireland }
\begin{document}
\section{Introduction}
Cohen-Kaplan-Katz-Unsal(CKKU)~\cite{Kaplan1},\cite{Kaplan2}
model is one of the promising formulations of the supersymmetric 
gauge theories without fine-tuning~\cite{others}. 
Their model is constructed from matrix model by 
using \textit{orbifolding}~\cite{Orbifold} and 
\textit{deconstruction}~\cite{deconstruction}, 
where the lattice spacing is dynamically obtained as 
the inverse vacuum expectation value of the scalar 
degrees of freedom in the matrix mode.

One possible problem in CKKU model is that the extended 
supersymmetry has flat directions for the scalar so that 
the lattice structure from the deconstruction suffers from 
the instability due to the quantum fluctuations of the scalar 
zero momentum modes. 
To suppress the divergence in the flat directions, 
soft susy-breaking terms for the scalar fields are introduced. 
Since such terms 
break the supersymmetry 
and cause the infra-red divergence, 
the original discussion of the renormalization based on 
exact supersymmetry on the lattice has to be modified 
by including the breaking terms. 
It is therefore important to re-examine the renormalization at 
1-loop level by explicit calculations in order to see whether this 
theory really needs fine-tuning or not.

\section{Method for perturbative calculation}
\label{Sec:method}
\subsection{Counter terms}
Before explaining our calculational method, let us 
discuss possible counter terms which needs renormalization.
Radiative corrections induce the operator $\mathcal{O}$ of the 
following structure into the action
\begin{equation}
\delta S=\frac{1}{g_2^2}Tr \int d^2z \mathcal{C_O} \mathcal{O}.
\end{equation}
Relevant or marginal operators ($\mathcal{O}$) whose canonical dimension 
$M[\mathcal{O}]=p$ 
at the $l$-loop correction must satisfy 
\begin{equation}  
p \le 4-2l 
\end{equation}
At 1-loop level, relevant or 
marginal 
operators with dimensions $0\le p \le 2$ 
can arise.  
At 2-loop level, relevant operators with the dimension $p=0$ 
can arise. Beyond 2-loop, there is no relevant 
or marginal counter term. 
Since the operator with the dimension $p=0$ is the cosmological constant, 
it does not play any serious 
role in fine-tuning problems.  

Let us now focus on the 1-loop relevant or 
marginal counter-terms.
Since bosonic fields have dimension 1 and fermionic fields have dimension 
$\frac{3}{2}$, the candidates for such operators are 
bosonic 1-point and 2-point functions.
Although fermionic 1-point functions are possible from dimension counting, 
they 
are forbidden by Grassman parity.

Since 1-point functions of gauge fields 
are forbidden from Furry's theorem and the 
2-point ones are also forbidden by the gauge symmetry. 
Hence the only possible 
counter terms are 
\begin{itemize}
\item $<s_x>,<s_y>$ (scalar 1point functions),
\item $<s^2_x>,<s^2_y>$ (scalar 2point functions).
\end{itemize}

In what follows, we will discuss 
the renormalization 
of these two operators.

\subsection{Calculational method for treating infra-red problems 
from fermion and bosons}

Let us explain the infra-red divergence problem from 
fermions and bosons.
It was pointed out by Giedt~\cite{Giedt} that there exists
an exact zero mode of the fermion matrix called 
^^ ever-existing zero mode'. Since it completely decouples 
from the action the fermion path-integral over this mode is 
ill-defined. Since there is no kinetic term for the 
zero momentum mode for the gauge field perturbative calculations 
based on the gaussian integral are also ill-defined.

In order to make the fermion path-integral well-defined
we propose to introduce the fermion mass term with coefficient 
$\mu_F$ inversely proportional to the lattice size $L$ 
to the action as
\begin{eqnarray}
S_2 & =& S_1 + \frac{a\mu_F \sqrt{2}}{g^2} Tr \sum_{\mathbf{n}} 
(\alpha_{\mathbf{n}} \bar{x}_{\mathbf{n}} \lambda_{\mathbf{n}}
+ \beta_{\mathbf{n}} \bar{y}_{\mathbf{n}} \lambda_{\mathbf{n}} 
-\alpha_{\mathbf{n}} y_{\mathbf{n+i}} \xi_{\mathbf{n}}
+ \beta_{\mathbf{n}} x_{\mathbf{n+j}} \xi_{\mathbf{n}}). 
\label{fermimass}
\end{eqnarray}
In order to make the path-integral over the zero momentum 
mode without kinetic terms we carry out non-perturbative 
calculation for the zero momentum modes while non-zero momentum 
modes are treated perturbatively.

Our calculational procedures ares as follows; 
\begin{enumerate}
\item Carry out the perturbative calculation only for non-zero 
      momentum modes, 
\item Carry out the non-perturbative calculation for zero-momentum modes, 
\item Take the large volume limit and investigate the behavior of green 
function and then it becomes clear whether fine-tuning is needed or not.
\end{enumerate}
For details see Ref.~\cite{Onogi:2005cz},

\section{Results}
\subsection{Non-zero mode contributions}\label{Sec:method2}
\subsubsection{1-point function}\label{1pper}
\begin{figure}
\begin{minipage}{.45\linewidth}
\includegraphics[width=5.8cm,angle=-90]{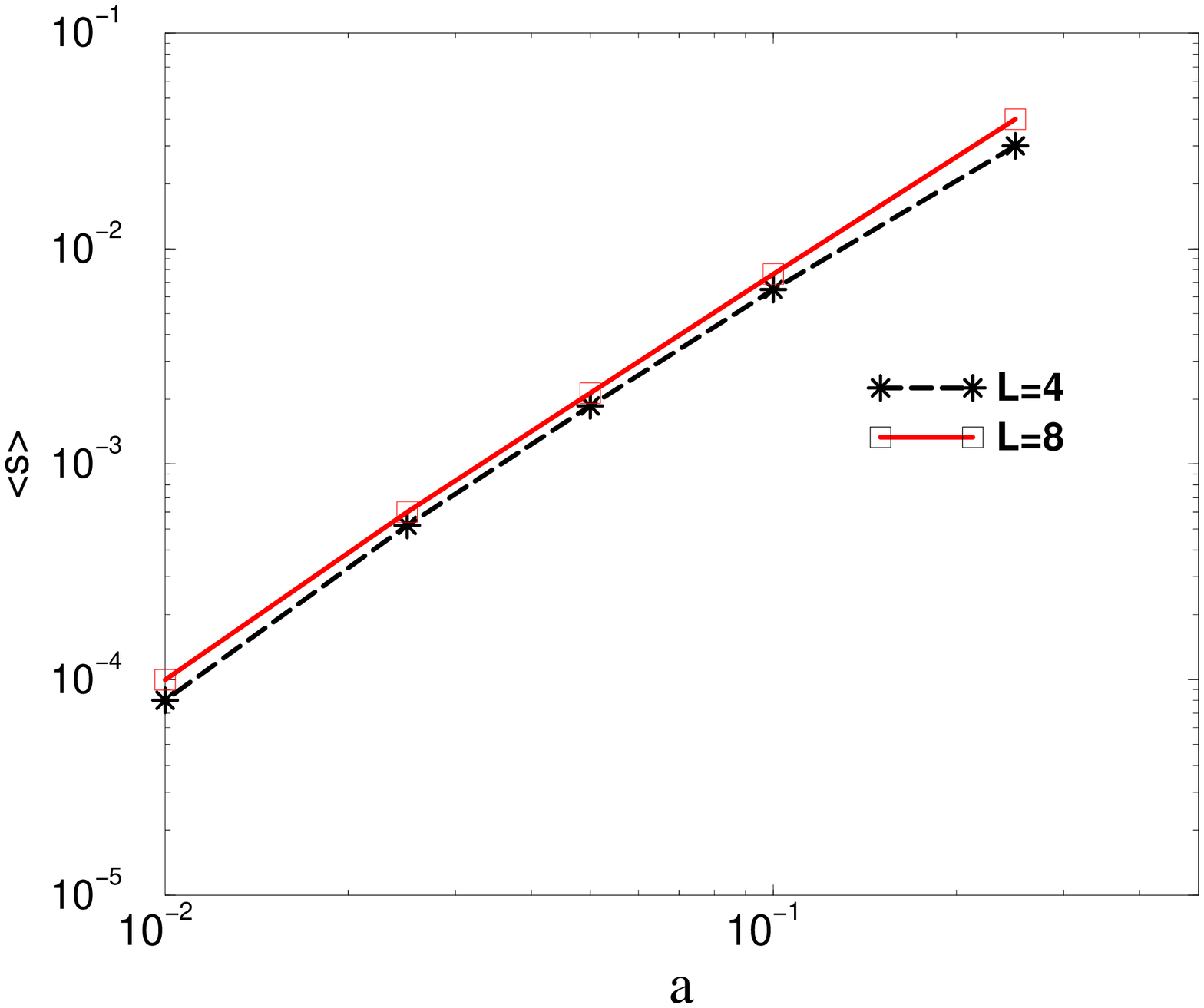}
\caption{$a$ dependence of the global minima
$\langle s \rangle$ of $V_{eff}$.
Horizontal axis is lattice spacing $a$, Vertical one is 
$\langle s \rangle$. 
We take here $g_2=1$, 
the solid line is for volume $L=8$, while the dashed line is for $L=4$.}
\label{hityou}
\end{minipage}
\begin{minipage}{.06\linewidth}
.
\end{minipage}
\begin{minipage}{.45\linewidth}
\includegraphics[angle=-90,width=7cm,clip]{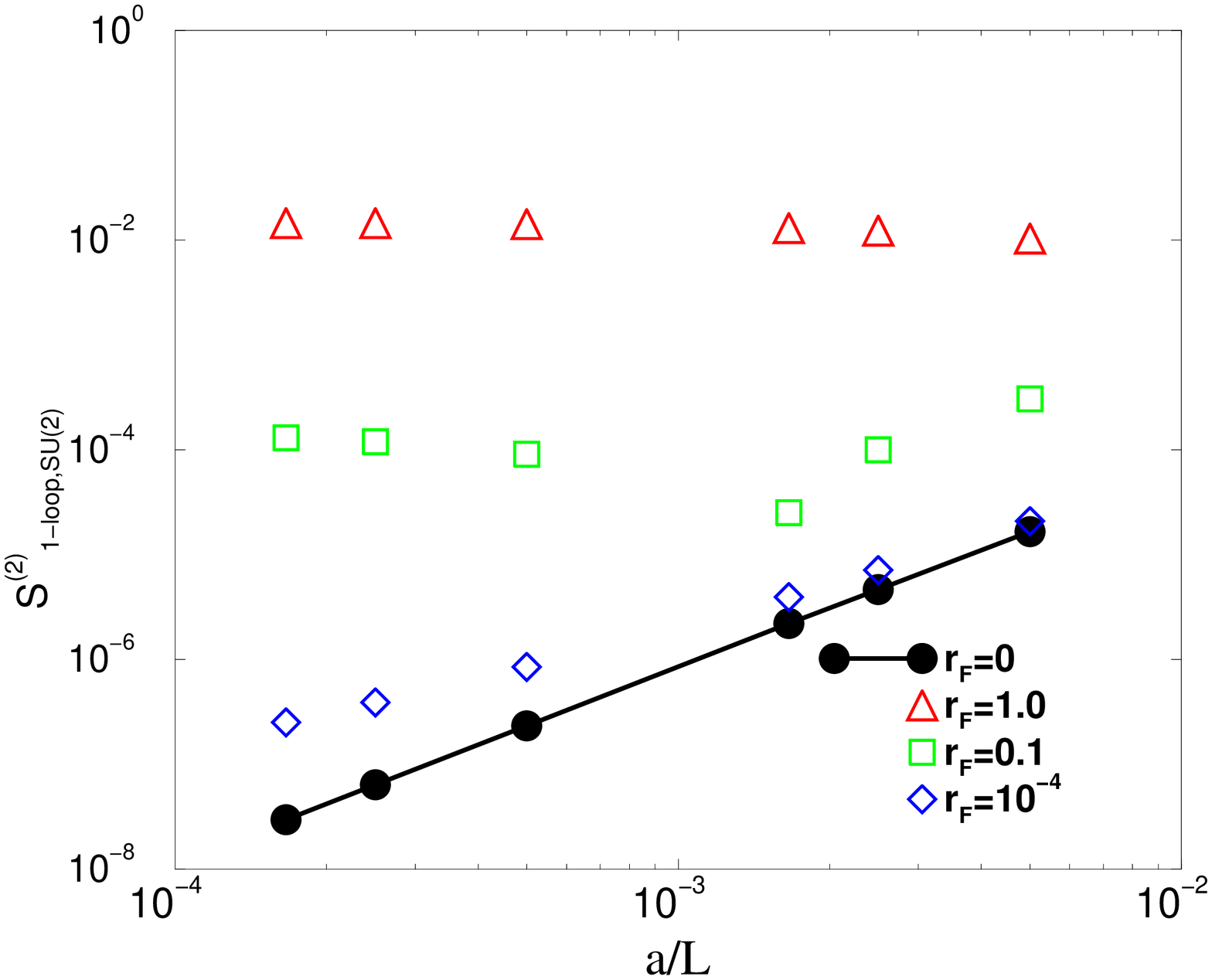}
\caption{($1/N$, $\bar{\mu}_F=r_F/N$) dependence of the nonabelian 
part of the 1-loop mass correction from the non-zero momentum mode. 
The horizontal axis is $\frac{1}{N}$ and the vertical axis is 
$S^{(2)}_{1-loop~SU(2)}$.}
\label{nonab}
\end{minipage}
\end{figure}

The only non-vanishing 1-point functions 
are the $U(1)$ part of the scalar fields $\tilde{s}^{0}_{x,y}(0)$.
These 1- point functions can be absorbed into the shift of the 
lattice spacing $a$ 
like as $\frac{1+ \langle s \rangle }{\sqrt{2} a}$
in Eqs.~(IV.1) and (IV.2) of Ref.\cite{Onogi:2005cz},
where $\langle s \rangle$ 
corresponds to the shift of the VEV.
In sufficiently weak coupling region we find that 
$\langle s \rangle$ vanishes quadratically in $a$ towards the continuum 
limit as shown in Fig.~\ref{hityou}.

\subsubsection{2-point function}\label{2p}

We next study whether 
the contribution from the non-zero momentum 
mode integral to the 2-point functions 
\begin{eqnarray}
(S^{(2)}_{1-loop})^{\alpha_1\alpha_2}_{\mu\nu}& = 
[&\delta_{\mu,0}\delta_{\nu,0}+\delta_{\mu,2}\delta_{\nu,2}]
[2\delta^{\alpha_1,0}\delta^{\alpha_2,0}S^{(2)}_{1-loop,~U(1)} 
+2M\delta^{\alpha_1,\alpha_2}S^{(2)}_{1-loop,~SU(2)} ]
\label{ym}
\end{eqnarray}
are relevant or not in the continuum limit. 
The numerical 
results of $S^{(2)}_{1-loop,~SU(2)}$
for several values of 
($1/N=a/L$, $\bar{\mu}_F\equiv r_F /N=a\mu_F$) with $\bar{\mu}\equiv
a\mu=1/N$ 
are given 
in Fig.~\ref{nonab} with $r_F$ fixed. 
$S^{(2)}_{1-loop,~U(1)}$ behaves same way as $S^{(2)}_{1-loop,~SU(2)}$.

From Fig.~\ref{nonab}, we find that the 1-loop correction 
for $r_F\neq 0$ does not vanish in the continuum limit, while 
that for $r_F=0$ vanishes. 
We should avoid these corrections which are independent from the 
volume $L$.
It becomes clear that one should adopt the following procedure
in order to avoid the appearance of such counter terms;  
\begin{enumerate}
\item Compute physical quantities for fixed ($1/N$, $\bar{\mu}_F=r_F/N$),
\item  Take $\mu_F \to 0$ with fixed $1/N$ first , i.e $r_F \to 0$,
\item  Then take the continuum limit, i.e.  $1/N \to 0$ .
\end{enumerate}
This two-step limit can avoid the counter terms as can be seen from 
Fig.~\ref{nonab} and Eq.~(V.17) of Ref.\cite{Onogi:2005cz} 
and make any loop correction for effective action irrelevant.

\subsection{Zero mode contributions}
\label{nonp}
\begin{figure}
\includegraphics[width=6cm,angle=-90,clip]{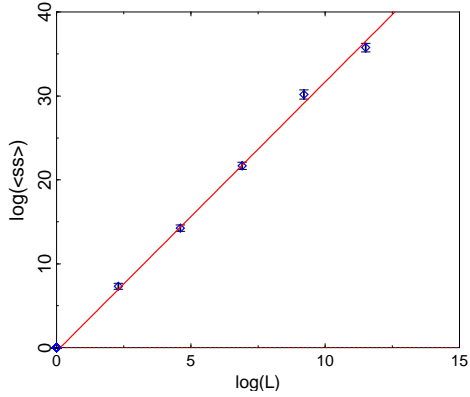}
\caption{The lattice size $L$ dependence of the 2-point function. 
The horizontal axis is $L$, 
where as the vertical axis is $\langle ss \rangle$. }
\label{0mode}
\end{figure}

We now carry out the nonperturbative integral 
over the zero momentum mode for the 1- and 2-point functions.
Since no term of 1-loop contributions 
from non-zero momentum modes 
to the effective action 
can survive in the continuum limit,
in order to evaluate 
1- and 2-point functions in the continuum limit,
we only have to perform the following integral
\begin{align}
I_n^{\alpha_1 \cdots \alpha_n}  = 
L^{\frac{3n}{2}}g_2^{\frac{n}{2}}
\frac{\int d\tilde{\phi}(\mathbf{0}) 
 det[D_{\psi}(\mathbf{0}) ]
\displaystyle{\prod_{i=1}^n}
(\tilde{s}^{\alpha_i}_{\mu}(0)) e^{-S_{fin}}}
{\int d\tilde{\phi}(\mathbf{0}) 
 det[D_{\psi}(\mathbf{0}) ]
e^{-S_{fin}}}, \label{patition3}\\
S_{fin}
 =  \sum_{\mu >\nu}Tr [\tilde{\phi}_{\mu}(0), \tilde{\phi}_{\nu}(0)]^2
  + \frac{\mu}{g_2}
Tr[ (\tilde{s}_x(0)^2
+ \tilde{s}_y(0)^2) ].
\label{sfin}
\end{align}
Among these integral, only 
the $SU(2)$ part of the 2-point function $I_2^{a,b}$ 
becomes relevant for the discussion whether fine-tuning is needed or
not. 
Since $I_2^{a,b}$ is the zero momentum mode of the propagator,
it can be written as $I_2^{a,b}=\delta^{a,b}L^2m_R^{-2}$ where $m_R^2$ is the 
renormalized mass squared. 
In order to numerically evaluate the 2-point function 
$I_2^{a,b}=\delta^{a,b}<ss>$, we carry out simulations 
in the  Metropolis algorithm with $2.0 \times 10^{5}$ sweeps 
for the thermalization and 
$2.0 \times 10^{7}$ sweeps for the measurement. 
We estimate the error by the variance with binsize of $100$ sweeps.
Since the 2-point function depends only on the product $g_2 L$, 
we take $g_2=1$ without loosing generality. 
Fig.~\ref{0mode} shows the $L$ dependence of the 
2-point function.
As can be seen in Fig.~\ref{0mode}, we find that 
$\langle ss \rangle$ increases with $L$.
Fitting the data with the following function
$\langle ss \rangle \sim  AL^{3+\alpha}$, 
we obtain $A=0.65(20)$ and $\alpha=0.210(46)$.
This gives the $L$ dependence of 
the renormalized mass 
\begin{equation}
m_R2\equiv \frac{\mu^2}{g_2^2}+\Delta \mu^2 \equiv L^2\langle ss 
\rangle^{-1} \sim  \frac{1}{AL^{1+\alpha}}, 
\label{mr}
\end{equation}
which vanishes in the large volume limit $L \to \infty$. 
Our result also implies that the contribution from the quantum 
corrections becomes dominant for large 
$L$.
Thus in the  continuum limit for finite volume, there is a 
non-trivial mass correction which is larger than the tree level contribution 
$\frac{\mu}{g_2}$. However, after taking the infinite volume limit 
the mass term vanishes so that there is no need for fine-tuning.

\section{Summary and conclusion}
We studied the CKKU model ~\cite{Kaplan1} 
for the supersymmetric U(2) gauge theory 
in two dimensions structure by an explicit perturbative calculation 
of scalar one-point and two-point functions. 
We pointed out that the naive perturbative calculation suffers from 
the infra-red divergences due to the flat directions in the zero 
momentum modes of gauge fields and fermion fields~\cite{Giedt}.
In order to avoid the infra-red divergence for the fermion zero mode, 
we introduce a new soft susy breaking mass term for the fermion
fields. For the bosonic fields, we apply the perturbation only for the
non-zero momentum mode and treat the zero momentum mode non-perturbatively. 
We found that there appears  non-trivial quantum mass corrections 
in the continuum limit.  However these corrections 
vanish in the infinite volume limit so that the CKKU model does not 
need fine-tuning to recover the full supersymmetry.
In addition to the fine-tuning problem discussed in this 
proceeding, 
several interesting results
are obtained by our explicit calculation as seen in Ref.~\cite{Onogi:2005cz}.
Firstly, 
we found the constraint for the parameter region 
where the lattice 
theory is well-defined. 
And secondly, it is found 
that the fermion-boson cancellation 
which suppresses the quantum corrections to the potential 
is needed 
to stabilize the deconstructed spacetime 
in the physical region where the lattice size is larger than the 
correlation length.
Similar instability has been observed in the non-perturbative 
study~\cite{Giedt} on the
bosonic part of the CKKU model for the (4,4) 2d
super-Yang-Mills~\cite{Kaplan2}.
For more details see Ref.~\cite{Onogi:2005cz}.

\end{document}